\documentclass[prc,aps,floats,floatfix,showpacs,nofootinbib,%
superscriptaddress]{revtex4}

\usepackage{amsmath}
\usepackage{amssymb}
\usepackage{amsfonts}
\usepackage{graphicx}
\usepackage{tabularx}
\usepackage{multirow}
\usepackage{lscape}
\usepackage{rotating}
\usepackage{pstricks}
\usepackage{feynmf}
\usepackage{lscape}
\usepackage{feynmf}
\begin{document}

\title{A new mass model for nuclear astrophysics: crossing 200 keV accuracy}

%%%%%%%%%%%%%%%%%%%%%%%%%%%%%%%%%%%%%%%%%%%%%%%%%%%%%%%%%%%%%%%%%%%%%%%%%%%%%%%

\author{M. Shelley}
\email{mges501@york.ac.uk}
\affiliation{Department of Physics, University of York, Heslington, York, Y010 5DD, UK}

\author{A. Pastore}
\email{alessandro.pastore@york.ac.uk}
\affiliation{Department of Physics, University of York, Heslington, York, Y010 5DD, UK}

%%%%%%%%%%%%%%%%%%%%%%%%%%%%%%%%%%%%%%%%%%%%%%%%%%%%%%%%%%%%%%%%%%%%%%%%%%%%%%%

 
\date{\today}

\begin{abstract}
By using a machine learning algorithm, we present an improved nuclear mass table with a root mean square deviation of less than $200$~keV. The model is equipped with statistical error bars in order to compare with available experimental data. We use the resulting model to predict the composition of the outer crust of a neutron star. By means of simple Monte-Carlo methods, we propagate the statistical uncertainties of the mass model to the equation of state of the system.
\end{abstract}

%%%%%%%%%%%%%%%%%%%%%%%%%%%%%%%%%%%%%%%%%%%%%%%%%%%%%%%%%%%%%%%%%%%%%%%%%%%%%%%

\maketitle

%%%%%%%%%%%%%%%%%%%%%%%%%%%%%%%%%%%%%%%%%%%%%%%%%%%%%%%%%%%%%%%%%%%%%%%%%%%%%%%

\section{Introduction}

Neutron stars (NS) are fascinating objects: with a typical mass of $M\approx1.5~M_\odot$ and radius~$R\approx12$~km~\cite{greif2019equation}, they represent the ideal laboratory to study the properties of nuclear matter under extreme conditions. Due to a strong pressure gradient, the matter within the NS arranges itself into layers with different properties~\cite{chamel2008physics}.
Going from the most external regions of the star to its centre, the matter density~$\rho$ spans several orders of magnitude from $\approx10^{-11}~\rho_0$ to $\approx3$--$5~\rho_0$, where $\rho_0=0.16\text{ fm}^{-3}\approx2.7\times 10^{14}\text{ g\,cm}^{-3}$ is the typical value of the density at the centre of an atomic nucleus~\cite{Book:Ring1980}.

The external region of a cold non-accreting NS is named the \emph{outer} crust. It consists of a Coulomb lattice of fully-ionised atoms with $Z$~protons and $N$~neutrons.
As discussed in Refs.~\cite{ruster2006outer, chamel2020analytical}, at $\beta$-equilibrium the composition of each layer of the crust at a given pressure~$P$ is obtained by minimising the Gibbs free energy per nucleon. The latter is the sum of three main contributions: the \emph{nuclear}, \emph{electronic} and \emph{lattice}. The effects of considering the system at a finite temperature have been presented in Ref.\cite{fantina2020crystallization}.
Since a large fraction of nuclei present in the outer crust are extremely neutron-rich, their binding energies are not known experimentally, and consequently one has to rely on a nuclear mass model. 
We refer the reader to Ref.~\cite{RevModPhys.89.015007} for a review of the properties of various equation of state (EoS) used to describe dense stellar matter.

Several models are available within the scientific literature with a typical accuracy, i.e., the root mean square~(RMS) deviation of the residuals, of $500$~keV~\cite{sobiczewski2018detailed}. In recent years, some of these mass models have been equipped with additional algorithms such as kernel ridge regression~\cite{wu2020predicting} or radial basis function interpolation~\cite{wang2011nuclear, niu2018high}, thus reducing the typical RMS to $\approx\text{200--300~keV}$.
Although such an RMS is remarkably low compared to the typical binding energy of a nucleus, the discrepancies between various models are still important, especially when used to predict the composition of the outer crust of a NS~\cite{wolf2013plumbing}.

Analysis of the residuals of various mass models shows that they do not show chaotic behaviour~\cite{barea2005nuclear}, thus it should be possible to further improve their accuracy, at least up to the level of Garvey-Kelson relations~\cite{garvey1966new}, by adding additional terms to account for the missing physics. This may be a very complex task, but machine learning methods can provide major support in achieving this goal.

In recent years, several authors have tried to reduce the discrepancy between theory and experiment by supplementing various mass models with neural networks~(NNs)~\cite{clark1999neural, athanassopoulos2004nuclear, athanassopoulos2005nuclear, utama2016nuclear, neufcourt2018bayesian, pastore2020impact}, where the NN learns the behaviour of the residuals. NNs are excellent interpolators~\cite{LESHNO1993861}, but they should be used with great care for extrapolation. The major problem is the presence of an unwanted trend related to the particular choice of the activation function. See Refs.~\cite{xu2020neural, pastore2020extrapolating} for a more detailed discussion on the topic.

A possible alternative to NNs has been discussed in Ref.~\cite{neufcourt2018bayesian}, and it is based on Gaussian processes~(GPs)~\cite{bastos2009diagnostics, pastore2017new, shelley2019advanced}. This GP method assumes that the residuals originate from some multivariate Gaussian distribution, whose covariance matrix contains some parameters to be adjusted in order to maximise the likelihood for the GP's fit to the residuals.
%The main advantage of a GP over a NN is that it conserves the mean value of the data: in this case by adjusting the GP parameters on the residuals, the GP will produce fluctuations around zero, thus conserving the trend.
The main advantage of a GP over a NN is that its predictions do not contain unwanted trends in extrapolation, but instead will always return to $0$ after a predictable extrapolation distance.
Moreover, GP predictions come equipped naturally with error bars. This is not the case for a standard NN (only Bayesian neural networks are equipped with posterior distributions that can be interpreted as error bars~\cite{neal2012bayesian}), and a more involved procedure is required to obtain an estimate~\cite{pastore2020extrapolating}.

In the current article, we present a new mass table, made by combining the predictions of a Duflo-Zucker~\cite{duflo1995microscopic} mass model with a GP, in order to further reduce the RMS of the residuals. We use the resulting model to analyse the composition of the outer crust of a NS. 
As previously done in Ref.~\cite{pastore2020impact}, we perform a full error analysis of the mass model and we use a Monte-Carlo procedure to propagate these statistical uncertainties through to the final EoS.

The article is organised as follows: in Sec.~\ref{sec:GP} we briefly introduce the concept of GPs and their use for regression, and in Sec.~\ref{sec:mass} we discuss the nuclear mass model and the improvement provided by the GP. In Sec.~\ref{sec:outer} we illustrate our results concerning the outer crust, and finally we present our conclusions in Sec.~\ref{sec:conclusion}.

%%%%%%%%%%%%%%%%%%%%%%%%%%%%%%%%%%%%%%%%%%%%%%%%%%%%%%%%%%%%%%%%%%%%%%%%%%%%%%%

\section{Gaussian process regression}\label{sec:GP}

\begin{figure}
        \centering
        \includegraphics{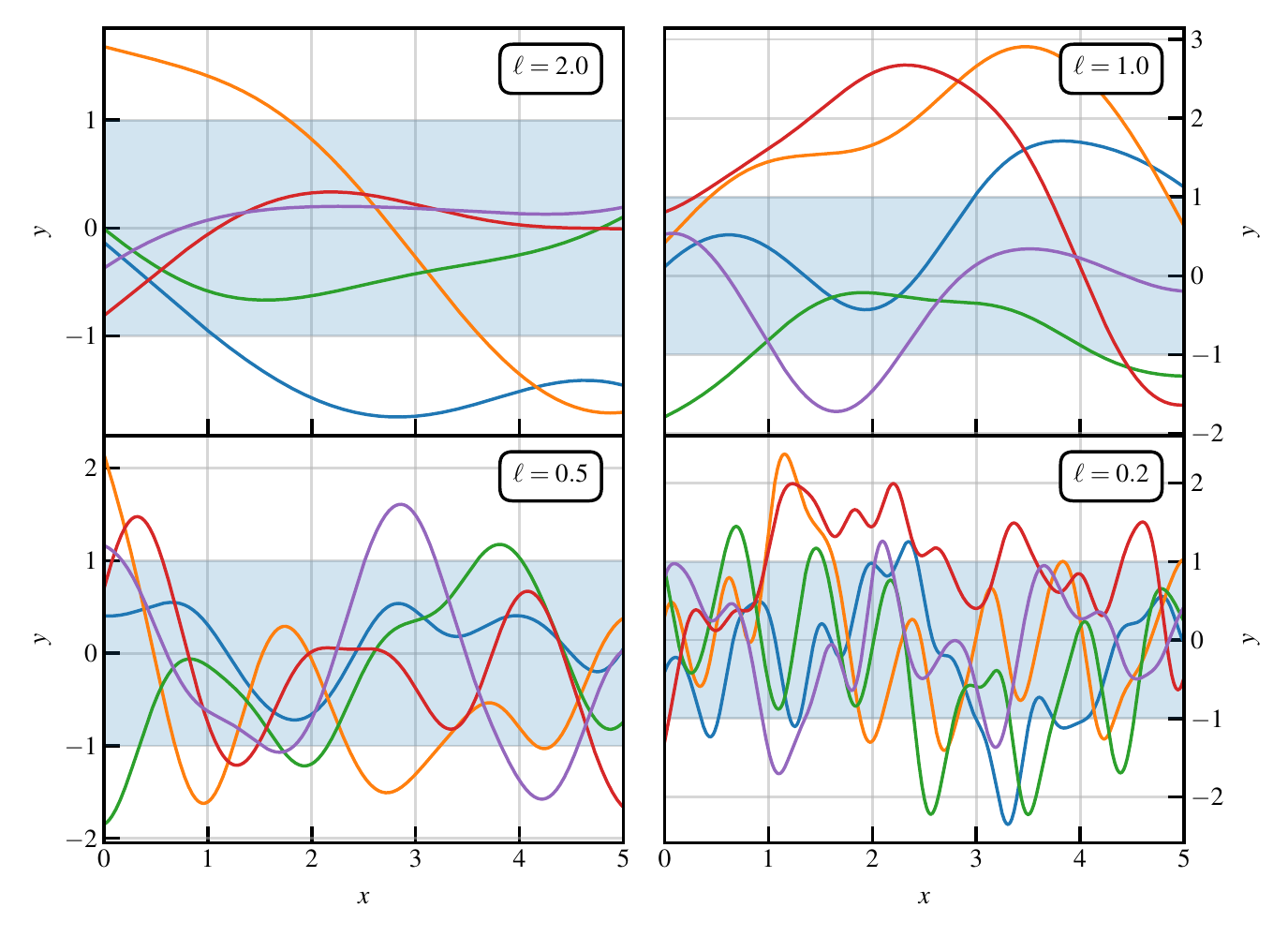}
        \caption{Colors online. Examples of the structure of \emph{prior} functions for various choices of the $\ell$ parameter. The shaded area represents the $1\sigma$ confidence interval.}
        \label{fig:GPprior}
    \end{figure}

\begin{figure}
        \centering
        \includegraphics{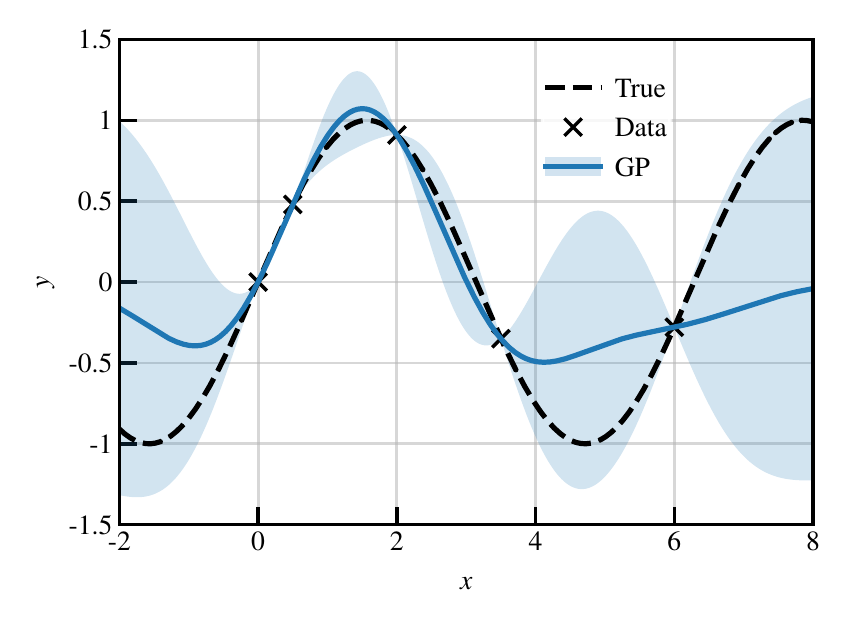}
        \caption{Demonstration of Gaussian process regression. The true function is $y=\sin(x)$, and the data points are at $x=\{0,0.5,2,3.5,6\}$. The solid line represents the GP mean, and the shaded areas give the $2\sigma$ confidence intervals. The optimised kernel parameters are $\eta^2=0.602,~\ell=1.063$. See text for details.}
        \label{fig:GPdemo}
    \end{figure}

We now introduce Gaussian processes, and their use as a regression tool. A Jupyter notebook is available as Supplementary Material; it was used to create Figs.~\ref{fig:GPprior} and~\ref{fig:GPdemo}, and contains additional plots which give a step-by-step introduction.

A Gaussian process~(GP) is an infinite-dimensional Gaussian distribution. Similar to how a one dimensional~(1D) Gaussian distribution has a mean~$\mu$ and variance~$\sigma^2$, a GP has a mean function~$\mu(\textbf{x})$, and a covariance function~$k(\textbf{x},\textbf{x}')$, also known as the kernel. In principle, $\textbf{x}$ can be a vector of length~$d$ representing a point in a $d$-dimensional input space, but for now we will just consider the case $d=1$, i.e., where $x$ is a single number. Just as we can draw random samples (numbers) from a 1D Gaussian distribution, we can also draw random samples from a GP, which are functions~$f(x)$. The kernel~$k(x,x')$ tells us the typical correlation between the value of $f$ at any two inputs $x$ and $x'$, and entirely determines the behaviour of the GP (relative to the mean function). For simplicity, we use here a constant mean function of $0$.

GPs can be used for regression of data if the underlying process generating the data is smooth and continuous. See Ref.~\cite{rasmussenGaussianProcessesMachine2006} for a thorough introduction to GPs for regression and machine learning. Many software packages are available for GP regression; in the current article we use the Python package \emph{GPy}~\cite{gpy2014}. For a set of data $\mathcal{Y}(x)=\{y_1(x_1),y_2(x_2),\dots y_n(x_n)\}$, instead of assuming a fixed functional form for the interpolating function, we treat the data as originating from a Gaussian process~$\mathcal{GP}$:

    \begin{eqnarray}\label{eq:gpe}
        \mathcal{Y}(x)\sim\mathcal{GP}(\mu(x),k(x,x')).
    \end{eqnarray}

\noindent No parametric assumption is made about the shape of the interpolating function, making GPs a very flexible tool. We adopt the commonly used \emph{RBF}~(radial basis function) kernel, also known as the squared exponential or Gaussian, which yields very smooth samples~$f(x)$, and has the form

    \begin{eqnarray}\label{eq:kernel}
        k_{\text{RBF}}(x,x') = \eta^2 \text{exp}\left[-\frac{\left(x-x'\right)^2}{2\ell^2}\right],
    \end{eqnarray}

\noindent where $\eta^2,\ell$ are parameters to be optimised for a given $\mathcal{Y}$. Both have easily interpretable meanings: $\eta$ gives the typical magnitude of the oscillations of $f(x)$, and $\ell$ the typical correlation length in $x$. When $\left|x-x'\right|$ is small, the correlation is large, and we expect $f(x)$ and $f(x')$ to have similar values. As $\left|x-x'\right|$ grows beyond a few correlation lengths~$\ell$, the correlation between $f(x)$ and $f(x')$ drops rapidly to $0$.

A simple way to understand GP is to make use of Bayes' theorem. Before doing the \emph{experiments} we have a prior distribution of $f(x)$, characterised by the kernel given in Eq.~\ref{eq:kernel}.
We can then draw sample functions from this prior, which are fully determined by the parameters $\eta^2,\ell$. In Fig.~\ref{fig:GPprior}, we show five sample draws of functions~$f(x)$ from some priors, which have $\eta=1$ and various choices of~$\ell$.
We observe that by varying $\ell$ we can have very different shapes in the prior samples. On average, they all lie within the shaded area representing $1-\sigma$ confidence interval 68\% of the time.

In Fig.~\ref{fig:GPdemo} we show a simple demonstration of GP regression, where the underlying true  function generating the data (dotted line) is simply $y=\sin(x)$.
We perform the \emph{experiment} and we extract five data points, indicated by crosses on the figure. 
The GP is fully characterised by two kernel parameters; clearly some sets of these parameters lead to better regression. For example, if $\ell$ is smaller than the typical data spacing, the GP mean will approach 0 in between data points, making it useless for interpolation (over-fitting); if $\eta^2$ is too large, the size of the confidence intervals will be overestimated. 
These parameters are determined using likelihood maximisation as discussed in  Ref.~\cite{gration2019dynamical}.

The \emph{GP mean} (solid line) here represents the average of all possible samples (from the \emph{posterior} distribution of $f(x)$) passing through the data~$\mathcal{Y}$ (crosses), i.e. the mean prediction. Since both the likelihood and the prior are Gaussian, so is the posterior. 
The GP mean is smooth, and interpolates all data points exactly. Outside the input domain, it approaches 0. As we would expect, the quality of the GP regressions is greatest where there is more data available, in this case ${0\leq x\leq4}$.

Also shown in Fig.~\ref{fig:GPdemo} are confidence intervals, here representing $2\sigma$~($\approx95\%$). The confidence intervals are 0 at each data point, and grow in between data points, more rapidly so when data are further apart. At the edges of the input domain, they also grow rapidly, representing the uncertainty in extrapolation, until reaching a maximum of~$2\eta$.
A very important aspect of the GP is the confidence intervals: in this case, we see that the \emph{true} function does not always match the GP mean, but $\approx95$\% of the true function falls within the $2\sigma$ interval.

%%%%%%%%%%%%%%%%%%%%%%%%%%%%%%%%%%%%%%%%%%%%%%%%%%%%%%%%%%%%%%%%%%%%%%%%%%%%%%%

\section{Nuclear masses}\label{sec:mass}

Nuclear mass models are used to reproduce the nuclear binding energies of all known nuclei, $\approx3200$~\cite{wang2017ame2016}. Within the mass database we distinguish two types of data: nuclear masses that have been directly measured ($\approx2400$) and the extrapolated ones ($\approx750$). The latter are obtained by indirect mass measurements and we will use them to benchmark our extrapolations.

In the current article, we use the Duflo-Zucker mass model\cite{duflo1995microscopic}; it consists of 10 terms ({DZ10} model), and is able to reproduce all known masses with a root mean square deviation of $\sigma_{\text{RMS}}\approx0.6$~MeV~\cite{pastore2020impact}. We refer the reader to Refs.~\cite{zuker2011anatomy, qi2015theoretical} for a detailed discussion on the different terms in the models.

The parameters of the DZ10 model have been adjusted in Ref.~\cite{pastore2020impact} using the block-bootstrap~(BB) method~\cite{pastore2019introduction}, yielding the optimal parameter set~$\mathbf{a}^0$. The reason for using BB is that it provides robust error bars on the parameters that take into account correlations between them~\cite{lahiri1999theoretical, bertsch2017estimating}.

\begin{figure}
        \centering
        \includegraphics{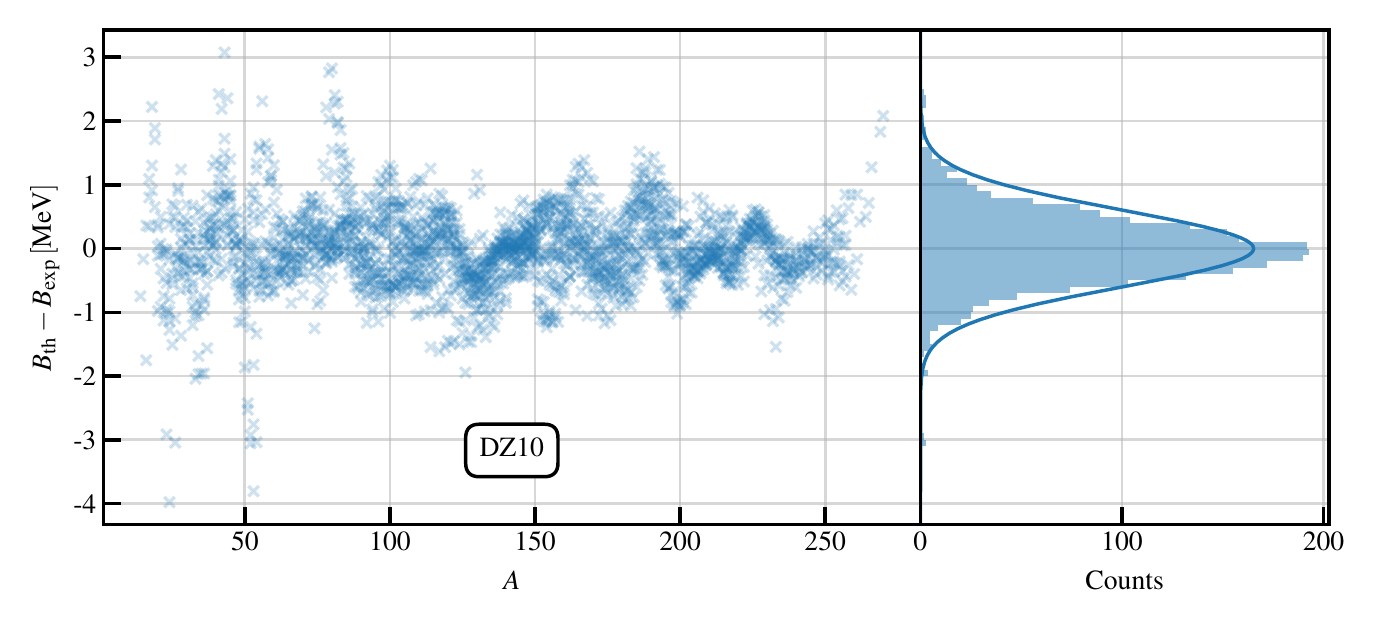}
        \caption{Left panel: residuals as a function of nucleon number~$A$ for the DZ10 model, for measured masses. In the right panel are the same residuals shown as a histogram, with a Gaussian fit overlaid (for which the mean is fixed to $0$, and the standard deviation to that of the residuals). See text for details.}
        \label{fig:residualsA_DZ10}
    \end{figure}

The assumption used to fit DZ10, as with any other mass model, is that the experimental binding energies~$B_{\text{exp}}(N,Z)$ are equal to the theoretical ones~$B_{\text{th}}(N,Z|\mathbf{a}^0)$ up to a Gaussian error~$\varepsilon(N,Z)$:
 
    \begin{eqnarray}\label{eq:hyp}
        B_{\text{exp}}(N,Z)=B_{\text{th}}(N,Z|\mathbf{a}^0)+\varepsilon(N,Z),
    \end{eqnarray}

\noindent where $B_{\text{th}}(N,Z)$ is the binding energy calculated using DZ10. In Fig.~\ref{fig:residualsA_DZ10}, we illustrate the residuals for DZ10 as a function of the nucleon number~$A=N+Z$. One clearly sees that these residuals show structure, thus indicating the presence of some missing \emph{physics} that is not properly accounted for by the model. In the right panel of the same figure, we plot the same residuals as a histogram, and we draw a Gaussian with mean~$0$ and width fixed to the RMS of the residuals. The height of the Gaussian is fitted on the residuals. We observe that the residuals display a Gaussian distribution.

A more detailed statistical test can be performed on these residuals to verify that they do not follow a regular Gaussian distribution --- see for example Refs.~\cite{carnini2020trees, pastore2020impact} for more details --- but for the current discussion a qualitative analysis is sufficient.

Having identified that there is room to improve the accuracy of the model, the most natural option to take is to add new terms~\cite{qi2015theoretical}. For example, a version of the Duflo-Zucker model with 33 parameters is available. Although the RMS reduces to $\approx300$~keV, the extra terms appear poorly constrained~\cite{qi2015theoretical}, and therefore the model is unsuitable for extrapolation.
We refer the reader to Ref.~\cite{nikvsic2016sloppy} for a detailed discussion on poorly constrained parameters.

Instead of explicitly \emph{creating} new terms for a given mass model, we can take advantage of machine learning methods. For example, in Refs.~\cite{utama2016nuclear, pastore2020impact}, the authors have adjusted a NN on the residuals of the DZ10 model in order to reduce the discrepancy between theory and experiment. The NN is able to reduce this discrepancy to a typical RMS of $\approx350$~keV~\cite{pastore2020impact}.

NNs are often very complex models, with several hundred free parameters. As discussed in~\cite{neufcourt2018bayesian}, a Gaussian process represents a valid alternative to a NN; the main advantages are the very small number of adjustable parameters, as discussed in Sec.~\ref{sec:GP}, and the superior performance on the database of nuclear masses when compared with a NN~\cite{neufcourt2018bayesian}.

\subsection{Augmenting the DZ10 model with a GP}

Having introduced the GP in Sec.~\ref{sec:GP}, we now apply it to the case of nuclear masses.
As done in Ref.~\cite{neufcourt2018bayesian}, we consider the same kernel given in Eq.~\ref{eq:gpe}, but now in the 2D case, meaning there are now three adjustable parameters. We also use a fourth parameter~$\sigma_n$, named the \emph{nugget}. The use of the nugget carries several advantages, including numerical stability~\cite{nealMonteCarloImplementation1997a}, and improved predictions~\cite{gramacyCasesNuggetModeling2012}. The kernel we use is then given by

    \begin{eqnarray}\label{eq:kernel2}
        k_{\text{RBF}}(x,x') = \eta^2 e^{-\frac{(N-N')^2}{2\rho_N^2}-\frac{(Z-Z')^2}{2\rho_Z^2}} + \sigma_n^2\delta_{xx'},
    \end{eqnarray}

where in the present case $x=(N,Z)$, and $\eta^2,\rho_Z,\rho_N$ are the adjustable parameters. Following Ref.~\cite{neufcourt2018bayesian}, $\rho_N$ and $\rho_Z$ are interpreted as correlation lengths in the neutron and proton directions, while $\eta^2$ gives the strength of the correlation between neighbouring nuclei.

The addition of the nugget means that the GP mean now does not necessarily pass directly through each data point, and that the $1\sigma$ confidence intervals only shrink to a minimum of~$\sigma_n$. After performing preliminary investigation using a full minimisation with all four parameters, we have found that the optimal value is $\sigma_n=0.2$~MeV. We have decided to fix this value, in order to simplify the analysis of the posterior distribution.
 
The main role of the nugget is to avoid over-fitting, which manifests itself via a correlation length smaller than the typical separation of the data. For example, setting $\sigma_n=0$~MeV would lead to a perfect reproduction of the data, but the resulting model would be totally useless; it would not be able to perform any kind of prediction, since the correlation lengths would be smaller than one (i.e., the separation the nuclear mass data). The nugget gives us an extra flexibility in identifying the residual correlations between the data as discussed in Ref.~\cite{pastore2020extrapolating}. 
For a more detailed discussion on GP and the role of the nugget we refer to Ref.~\cite{neufcourt2018bayesian}.

\begin{figure}
        \centering
        \includegraphics{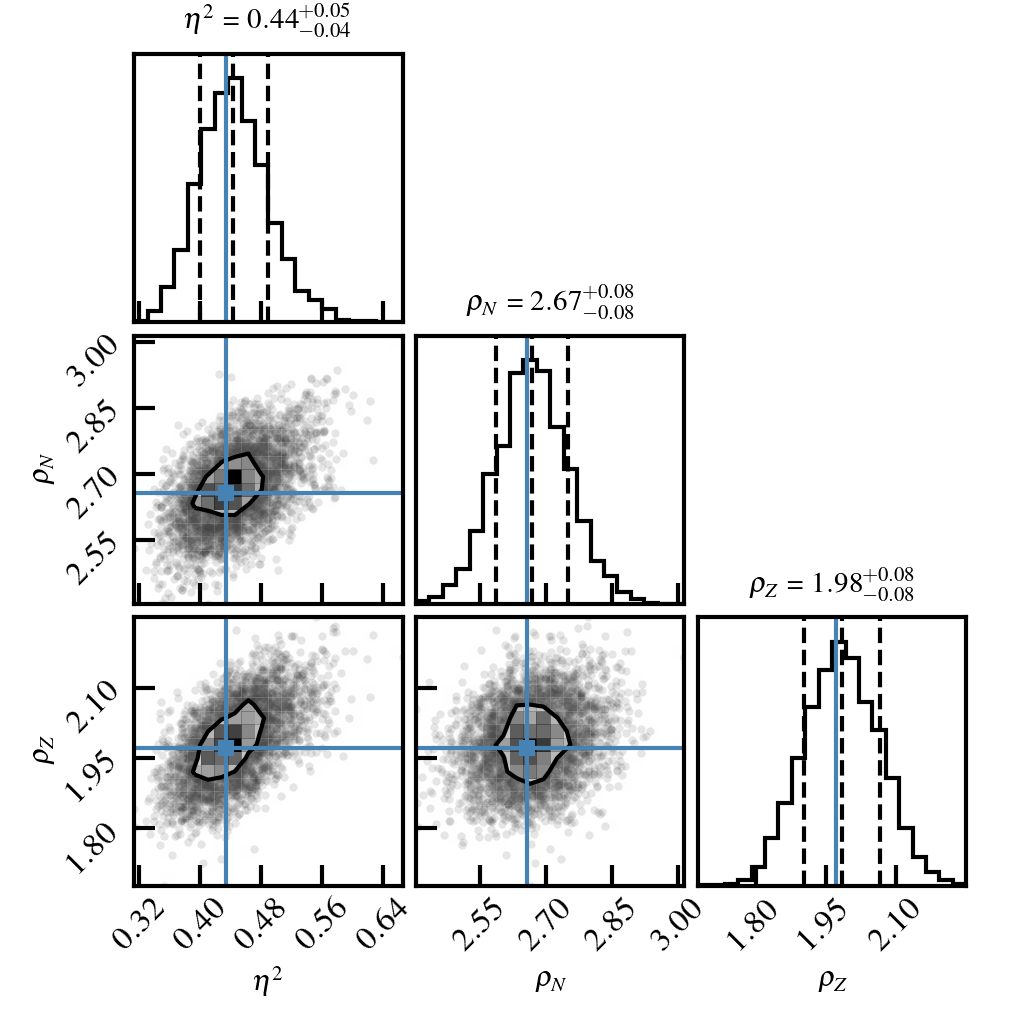}
        \caption{Posterior distributions of GP parameters obtained through MCMC sampling. The horizontal and vertical solid lines indicate the optimal parameter values obtained by maximising the likelihood. The vertical dotted lines on each 1D histogram indicate the mean and $1\sigma$ confidence intervals obtained through the MCMC sampling. See text for details.}
        \label{fig:GP_corner}
    \end{figure}

As discussed previously, we adjust the parameters of the GP on the residuals of the DZ10 model (shown in Fig.~\ref{fig:residualsA_DZ10}). The parameters $\eta,\rho_N,\rho_Z$ are determined through maximising the likelihood for the GP. See Ref.~\cite{gration2019dynamical} for details.
In Fig.~\ref{fig:GP_corner}, we illustrate the posterior distribution of the parameters in the form of a corner plot. The distributions were obtained with Markov Chain Monte-Carlo~(MCMC) sampling~\cite{geyer1992practical}. The plot illustrates the shapes of the distributions around the optimal parameter set, and it provides us with the error bars for the parameters and information about their correlations. In this case we see that all parameters are very well determined by the residuals data, and a weak correlation is observed between $\eta$ and $\rho_N$, and between $\eta$ and $\rho_Z$.

A very interesting result is that the two correlation lengths $\rho_{N,Z}$ are as large as, or greater than, $2$. This means that, if we know the residual for a nucleus with mass number~$A$, we can infer properties of the nucleus with $A\pm2$. This result is in agreement with the analysis done in Ref.~\cite{pastore2020impact}, which was based on the auto-correlation coefficients. 

\begin{figure}
        \centering
        \includegraphics{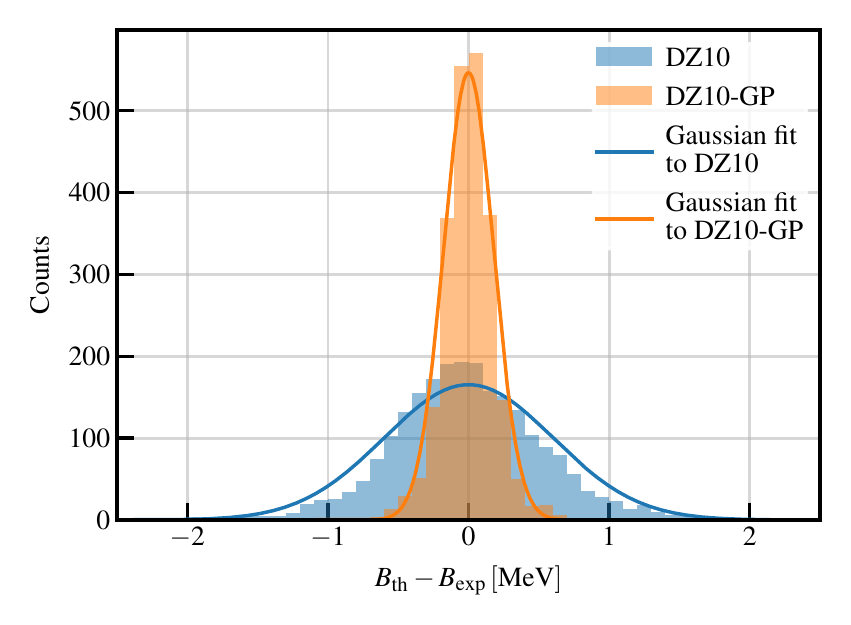}
        \caption{Distributions of the residuals for the DZ10 and DZ10-GP models, for measured masses. Gaussian fits to the residuals are also shown, with the mean fixed to $0$, and the standard deviation to that of the residuals. See text for details.}
        \label{fig:residualDistsMeasured}
    \end{figure}

We now construct our new model for $B_{\text{th}}$ (appearing in Eq.~\ref{eq:hyp}) as $B_{\text{th}}=B_{\text{DZ10}}-GP$, which we name \emph{DZ10-GP}. In Fig.~\ref{fig:residualDistsMeasured} we compare the residual distributions for the DZ10 and DZ10-GP models for measured masses. We see that the RMS of the DZ10 model has been greatly reduced. The RMS of the DZ10-GP model is $\sigma=178$~keV, which at the moment is probably among the lowest values ever obtained using a mass model fitted on all the available masses, with a total of $10+4=14$ adjustable parameters.

\begin{figure}
        \centering
        \includegraphics{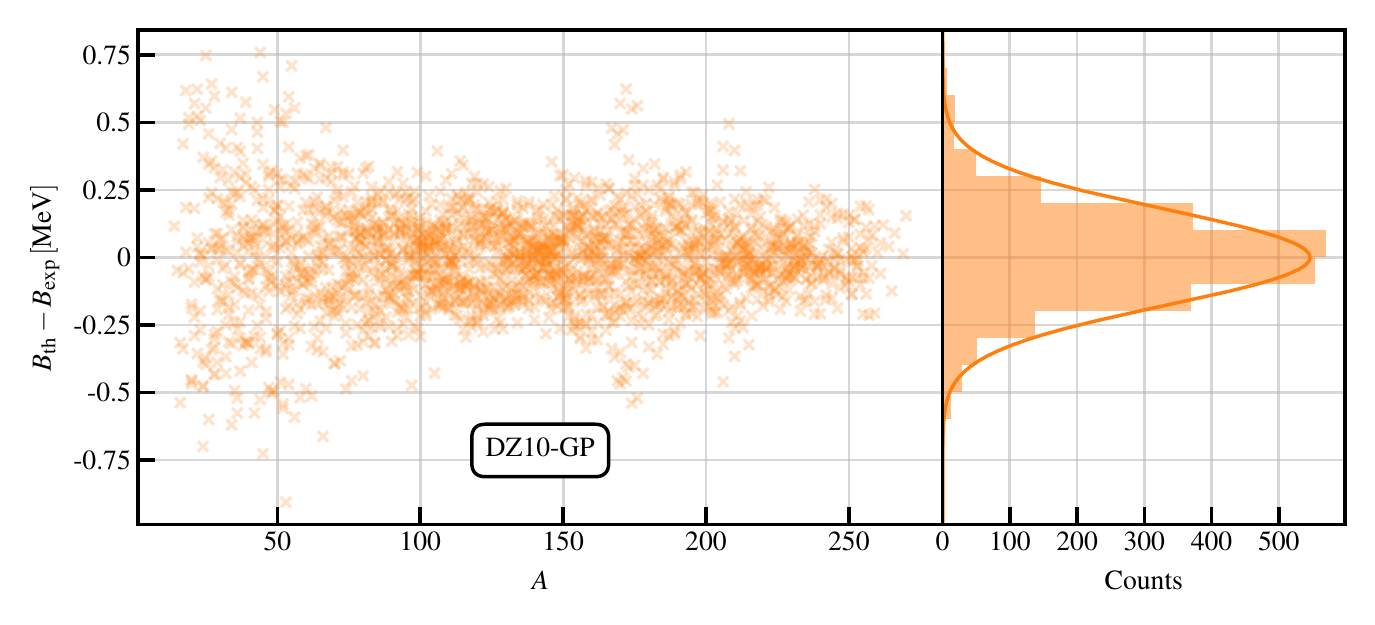}
        \caption{The same as Fig.~\ref{fig:residualsA_DZ10}, but for the DZ10-GP model. See text for details.}
        \label{fig:residualsA_DZ10-GP}
    \end{figure}

In Fig.~\ref{fig:residualsA_DZ10-GP}, we illustrate the residuals obtained from the DZ10-GP model as a function of mass number~$A$. We clearly see that the GP has been able to capture the missing physics of the DZ10 model, in particular smoothing out the spikes observed in Fig.~\ref{fig:residualsA_DZ10}. We observe that the maximum discrepancy between theory and experiment is now always lower than $1$~MeV, and the structure observed in Fig.~\ref{fig:residualsA_DZ10} has now disappeared, with the new residuals exhibiting behaviour close to white noise. The presence or not of white noise in the model may represent a lower bound on the accuracy one can achieve with a theoretical model, as discussed in Ref.~\cite{barea2005nuclear}; we leave such an interesting analysis for a future investigation.

\subsection{Extrapolation using the DZ10-GP model}

\begin{figure}
        \centering
        \includegraphics{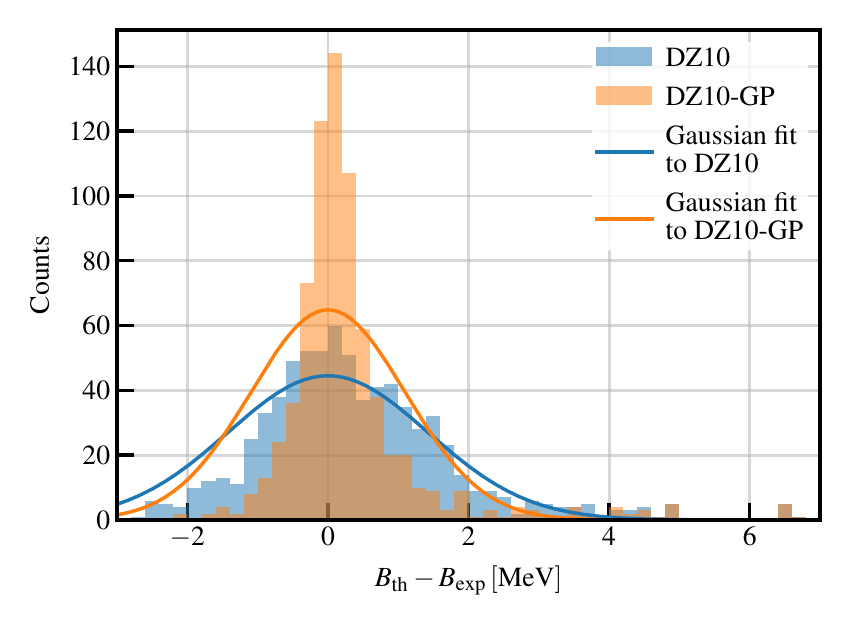}
        \caption{Same as Fig.~\ref{fig:residualDistsMeasured}, but for extrapolated masses. See text for details.}
        \label{fig:residualDistsExtrapolated}
    \end{figure}

Having created the DZ10-GP model, we now benchmark its extrapolations on the set of $\approx750$ nuclear masses obtained via indirect measurements~\cite{wang2017ame2016}. The results are presented in Fig.~\ref{fig:residualDistsExtrapolated}. The original DZ10 model gives an RMS of $1.426$~MeV; the inclusion of GP corrections reduces the RMS to $1.100$~MeV. It is worth noting that some outliers are still present. We have checked that the six nuclei with a residual larger than $6$~MeV are all in the region of super-heavy nuclei with $Z\ge108$.

\begin{figure}
        \centering
        \includegraphics{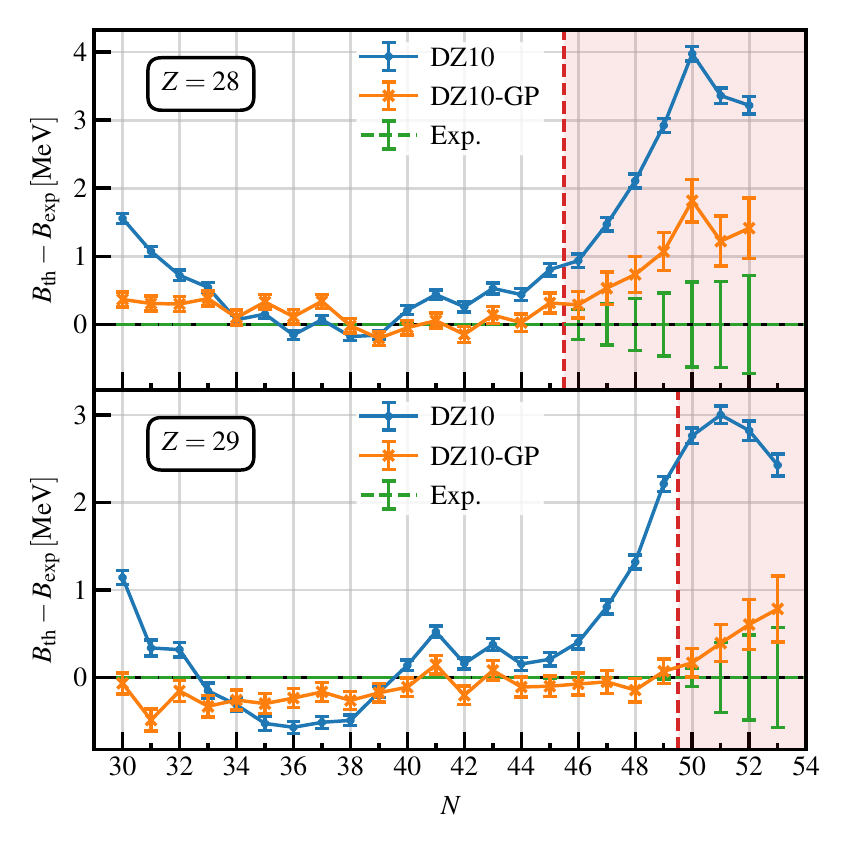}
        \caption{Residuals for the DZ10 and DZ10-GP models, for the $Z=28$ and $Z=29$ isotopic chains. The vertical dashed lines represent the transition from nuclei used for training to nuclei for which predictions are made. See text for details.}
        \label{fig:Zchains}
    \end{figure}

Since the main goal of this article is the study of the outer crust of a neutron star, in Fig.~\ref{fig:Zchains} we illustrate in great detail the evolution of the residuals for two isotopic chains --- copper and nickel --- that play a very important role in determining the composition of the outer crust~\cite{wolf2013plumbing}.

We observe that the original DZ10 model reproduces fairly well the data in the middle of the isotopic chains, and that it tends to give large discrepancies at the edges. Even the inclusion of the statistical error bars of DZ10 are not enough to explain such a discrepancy. We refer the reader to Ref.~\cite{pastore2020impact} for a detailed discussion on how these error bars have been obtained. 
On the contrary, the use of the GP helps to flatten out the discrepancies, and produces predictions very close to the data in the extrapolated region. By considering the experimental and the theoretical error bars, we observe that our DZ10-GP model reproduces these data reasonably well.
The error bars of the DZ-GP model have been obtained using a \emph{n\"aive} approach, i.e., summing in quadrature the statistical error bars of the original DZ model and the confidence intervals of the GP model. 

As done in Ref.~\cite{pastore2020impact}, we validate the error bars by comparing with experimental masses. In particular, we expect that 68\% of known masses differ from the model prediction no more than $\sigma=\sigma_{\text{th}}+\sigma_{\text{exp}}$, where $\sigma_{\text{th}}$ is the theoretical error bar of the DZ10-GP model and $\sigma_{\text{exp}}$ is the experimental error bar. By increasing the error bar by a factor of $2$ and $3$ we should obtain 95\% and 99.7\% of experimental binding energies falling into the interval. 

    \begin{table}[!h]
        \centering
        \begin{tabular}{c|ccc}
        \hline
        \hline
                    & $1\sigma$ & $2\sigma$ & $3\sigma$\\
                    \hline
         Full chart & 61\% & 88.8\% & 96.2\%\\
         $50\le A\le 150 $ & 59.2\% & 89.1\% & 97.3\%\\
              $20\le Z\le 50 $ & 54.4\% & 84.1\% & 95.5\%\\
        \hline
        \hline
        \end{tabular}
        \caption{Percentage of nuclei included in the total error bars for the DZ10-GP model for three different sectors of the nuclear chart}
        \label{tab:err}
    \end{table}

From Tab.~\ref{tab:err}, we observe that most of the nuclei fall within these error bars as expected, although we still underestimate in some relevant regions of the chart, such as ${20\le Z\le 50}$ which is important for outer crust calculations. This discrepancy may be a sign of other contributions to the error bar that were not taken into account here, for example correlations between the DZ10 and GP error bars.

\begin{figure}
        \centering
        \includegraphics{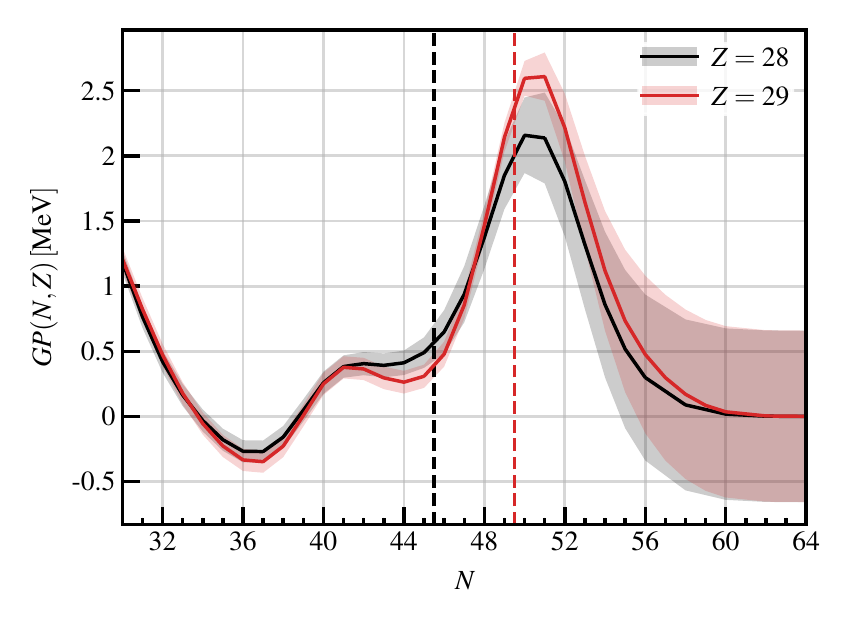}
        \caption{GP correction for $Z=28$ and $Z=29$. The vertical dashed lines represent the transition from nuclei used for training to nuclei for which predictions are made. The shaded ares represent the GP $1\sigma$ error bars. See text for details.}
        \label{fig:correction}
    \end{figure}

In Fig.~\ref{fig:correction}, we show the evolution, along two isotopic chains, of the GP's contribution to binding energy. We see that these contributions drop to $0$ as the neutron-rich region is approached.
On the same figure, we also report the evolution of a $1\sigma$ error bar provided by the GP. As discussed previously, we notice that the error bars grow towards the neutron drip-line, where we have little or no available data to constrain the GP.

From Fig.~\ref{fig:correction}, we observe that the confidence interval provided by the GP model at large values of N becomes constant and equal to~$\eta$. This means that at very large extrapolations, the GP error bar is most likely underestimating. In this case, the model error bar should become larger and be the dominant source of error. See for example~\cite{gao2013propagation}. 

This behaviour can be understood from the value of the GP's correlation length for neutrons, $\rho_N=2.67$: by construction the GP predictions tend to the mean of the data, in this case $0$, after $\approx2\text{-}3$~times~$\rho_N$. This means that the GP will be effective in describing extrapolated neutron-rich nuclei with at most $\approx8\text{-}10$~neutrons more than the last nucleus in our training set.
This is clearly only a rule of thumb, but it is enough to cover most of the extrapolated nuclei that are present in the outer crust~\cite{pearson2011properties} of a neutron star.
For nuclei further away from the known data-set, the extrapolation is governed by the underlying nuclear mass model, i.e., the DZ10 model.
This is not the case for other approaches, for example with NNs that can introduce an additional trend on top of the model. Such a trend is difficult to predict \emph{a priori}, and it may be strongly biased by the training method. See Ref.~\cite{pastore2020extrapolating} for a more detailed discussion.

\subsection{Comparison with AME2020}

Having trained and developed the DZ10-GP model on the AME2016 database~\cite{wang2017ame2016}, we now benchmark the predictions against the newly published AME2020 database~\cite{huang2021ame}. Between the 2016 and 2020 database, we have 74 new isotopes.

\begin{figure}
        \centering
        \includegraphics{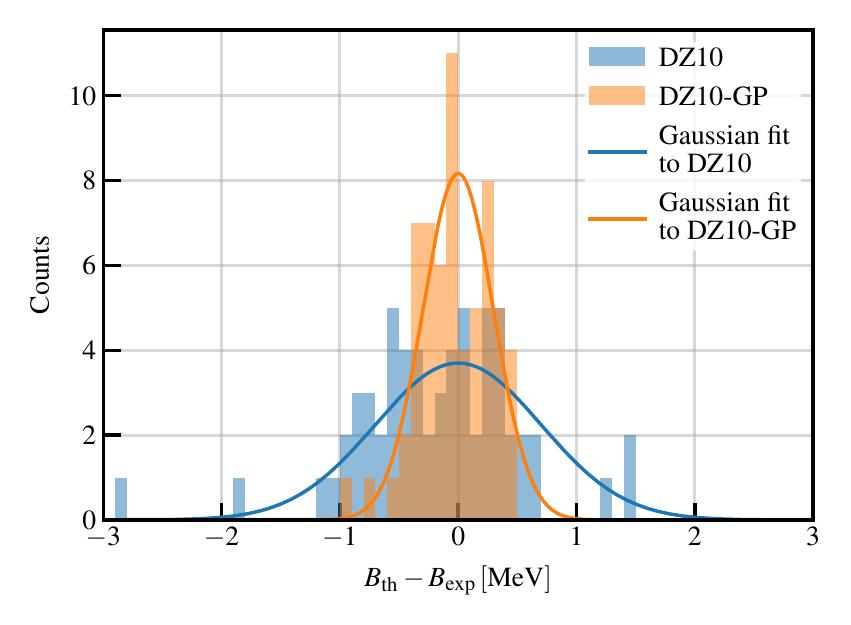}
        \caption{Distributions of the residuals for the DZ10 and DZ10-GP models, for new masses presented in AME2020~\cite{huang2021ame}. Gaussian fits to the residuals are also shown, with the mean fixed to $0$, and the standard deviation to that of the residuals. See text for details.}
        \label{fig:ame2020}
    \end{figure}

In Fig.~\ref{fig:ame2020}, we report the distribution of the residuals for the new isotopes presented in AME2020 database, apart from the Cu measurements already published in Ref.~\cite{welker2017binding}.
We observe that the RMS of the original DZ10 model for these new data is $\sigma_{\text{DZ10}}=701$~keV, while for the DZ10-GP model it is $\sigma_{\text{DZ10\text{-}GP}}=299$~keV. Notice that in this case we do not re-adjust the GP model over the new data.
This test clearly proves that the GP is not over-fitting the data, but it was really able to grasp a signal in the residuals and is therefore capable of performing extrapolations in regions in the proximity of the data set used for the training. We also observed that 50\% of the new isotopes fall within the error bars of the original DZ10-GP model. This value is slightly lower than what is reported in Tab.~\ref{tab:err}, but still reasonable compared to the expected 68\%.

%%%%%%%%%%%%%%%%%%%%%%%%%%%%%%%%%%%%%%%%%%%%%%%%%%%%%%%%%%%%%%%%%%%%%%%%%%%%%%%

\section{Outer crust}\label{sec:outer}

To determine the chemical composition of the outer crust, we minimise the Gibbs free energy per particle, which is defined as~\cite{baym1971ground}

    \begin{eqnarray}\label{eq:gibbs}
        g=\mathcal{E}_{nuc}(N,Z)+\mathcal{E}_{e}(A,Z)+\mathcal{E}_{l}(A,Z)+\frac{P}{\rho_b}\;,
    \end{eqnarray}

\noindent where $\rho_b$ is the baryonic density. The three terms $\mathcal{E}_{nuc},\mathcal{E}_{e},\mathcal{E}_{l}$ are the \emph{nuclear}, \emph{electronic} and \emph{lattice} energies per nucleon respectively~\cite{basilico2015outer}.
The pressure~$P$ arises only from lattice and electron contributions as $P=P_L+P_e$. For more details, we refer to Ref.~\cite{baym1971ground}, where the entire formalism has been discussed in great detail. 

The novelty of the current approach is in the treatment of the nuclear term, which takes the form 

    \begin{eqnarray}
        \mathcal{E}_{nuc}(N,Z)=\frac{Zm_p+Nm_n}{A}-\frac{\mathcal{B}(N,Z)}{A}
    \end{eqnarray}

\noindent where $m_{p(n)}$ is the mass of the proton (neutron) and $\mathcal{B}$ is the nuclear binding energy given by the mass model. In the current article, we use the mass model DZ10-GP as discussed in Sec.~\ref{sec:GP}.
The composition predicted by the mass models is given in Tab.~\ref{tab:DZcompo}. By comparing the DZ10-GP results with those obtained using only the DZ10 model, we observe some discrepancies in the extrapolated region at low~$P$. In particular, we notice that the improved mass model (DZ10-GP) predicts the existence of $^{80}$Zn, that is not considered in the original DZ10 model. At higher $P$, the two mass models give very similar results. This is simple to understand since, as discussed in Sec.~\ref{sec:GP}, the GP correction tends to $0$ for large extrapolations, as seen in Fig.~\ref{fig:correction}.

    \begin{table}[!h]
    \setlength{\tabcolsep}{10pt}
    \renewcommand{\arraystretch}{1.3} 
    \begin{center}
    \begin{tabular}{ccc|ccc}
    \hline
    \hline
    \multicolumn{3}{c}{DZ10} & \multicolumn{3}{c}{DZ10-GP}\\
    \hline
     $P_{max}$ [MeVfm$^{-3}$]  &N &Z & $P_{max}$ [MeVfm$^{-3}$]  &N &Z  \\
    \hline 
        3.30 $\cdot 10^{-10}$       &30 &26 &  3.30 $\cdot 10^{-10}$       &30 &26\\
       4.36$\cdot 10^{-8}$     &34 &28 & 4.36$\cdot 10^{-8}$     &34 &28 \\
     3.56$\cdot10^{-7}$     & 36   &       28&  3.56$\cdot10^{-7}$     & 36   &       28\\
        4.02$\cdot10^{-7}$          &    38       &   28& 4.02$\cdot10^{-7}$          &    38       &   28  \\
       1.03$\cdot10^{-6}$     &      50   &       36 &1.03$\cdot10^{-6}$     &      50   &       36  \\
       5.59$\cdot10^{-6}$       & 50      &    34  &  5.59$\cdot10^{-6}$       & 50      &    34   \\
                      \hline
       1.76$\cdot10^{-5}$  &          50     &     32  & 5.59  $\cdot10^{-6}$ & 50 & 32 \\
                    &    &          &  1.77$\cdot10^{-5}$ & 50 & 30  \\
             1.58$\cdot10^{-4}$    &   50 &         28 &  3.22$\cdot10^{-5}$ & 50 & 28  \\
       1.82$\cdot10^{-4}$    &    82        &  42   &  1.21$\cdot10^{-4}$    &    82        &  42\\
       3.31$\cdot10^{-4}$    &      82       &   40 &  1.81$\cdot10^{-4}$    &      82       &   40  \\
       4.83$\cdot10^{-4}$  &     82         & 38   &  3.31$\cdot10^{-4}$  &     82         & 38 \\
       4.86$\cdot10^{-4}$ &          82         & 36 &4.84$\cdot10^{-4}$ &          82         & 36  \\
          \hline
          \hline
    \end{tabular}
    \caption{Composition of the outer crust of a NS using the DZ10 and DZ10-GP mass models. In the first and fourth columns we report the maximum value of pressure at which the nucleus is found using the minimisation procedure. The horizontal line separates the measured and extrapolated masses reported in AME2016~\cite{wang2017ame2016}.}
    \label{tab:DZcompo}
    \end{center}
    \end{table}

Since our goal is to obtain the statistical uncertainties of the equation of state, we perform a simple Monte-Carlo sampling of the error bars of our DZ10-GP model (under a Gaussian assumption). We generate $10^4$ new mass tables, and we use them to calculate the composition of the outer crust.

Using a \emph{frequentist} approach~\cite{bar89}, we define the existence probability of each nucleus as the ratio of the number of times a given nucleus appears in the various EoS at a given pressure, divided by the total number of mass tables. See Ref.~\cite{pastore2020impact} for more details.

\begin{figure}
        \centering
        \includegraphics[width=0.5\textwidth,angle=0]{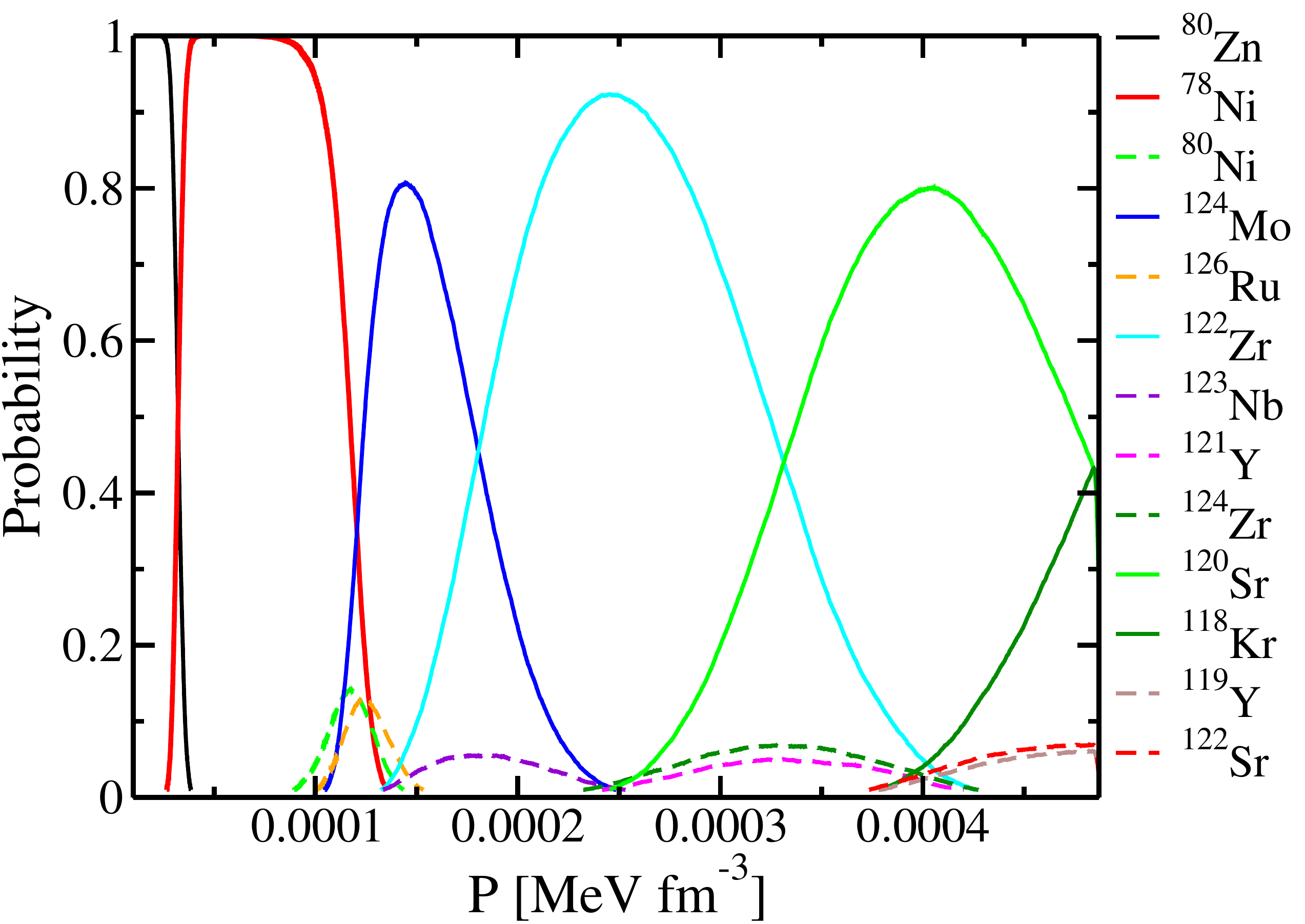}
        \caption{Colors online. Existence probability of a given nucleus within the outer crust as a function of the pressure, obtained via a Monte-Carlo sampling using the DZ10-GP mass table. See text for details.}
        \label{fig:prob}
    \end{figure}

In Fig.~\ref{fig:prob}, we show the evolution of the existence probability for each nucleus in the outer crust as a function of the pressure of the star.
We notice that, as confirmed by other authors~\cite{pearson2011properties}, the favourable configurations are those close to the neutron shell closures at $\text{N}=50$ and $\text{N}=82$. However, due to the large error bars, there is a non-negligible probability for several nuclei to be present within the outer crust.

\begin{figure}
        \centering
        \includegraphics[width=0.5\textwidth,angle=0]{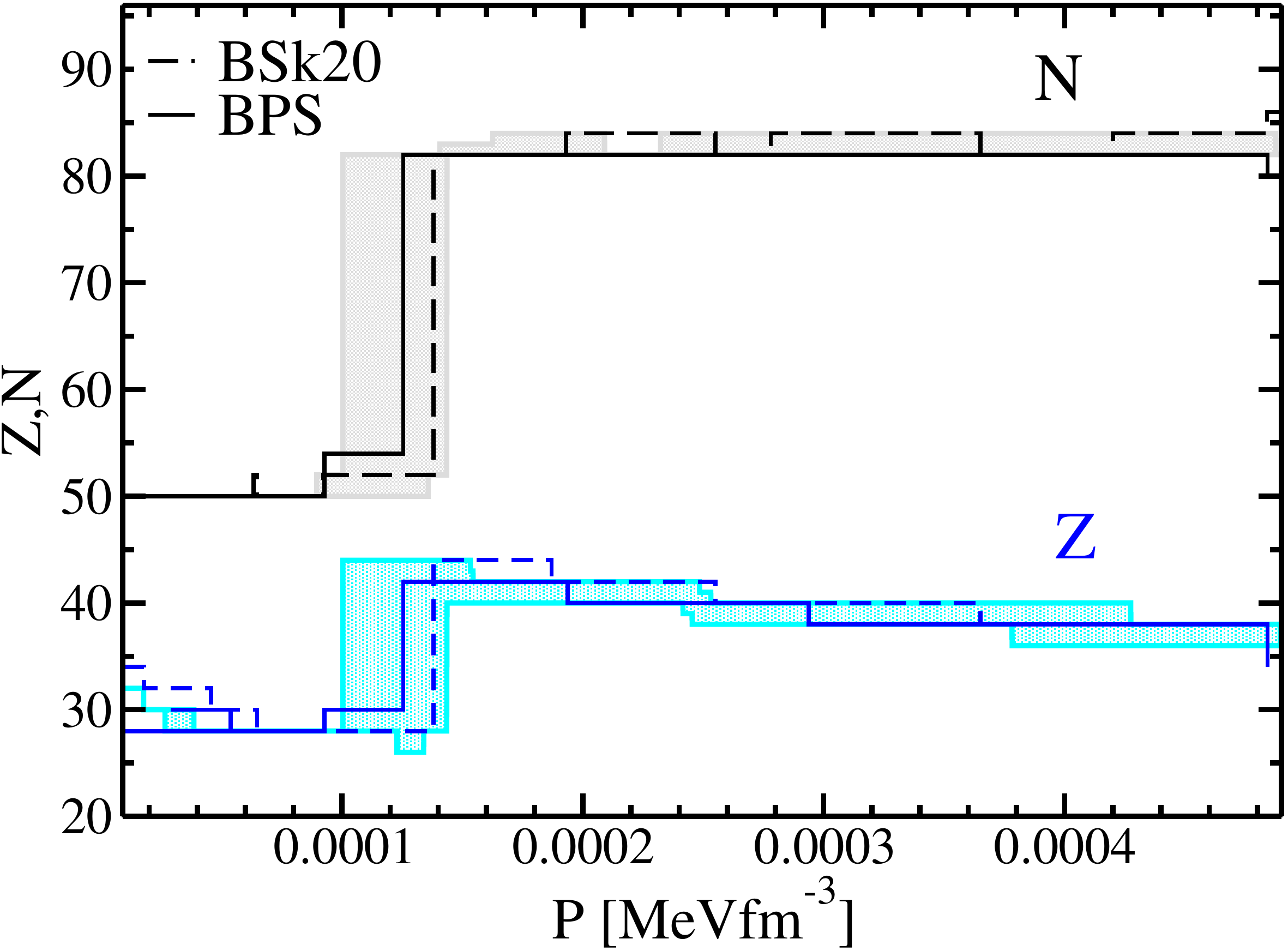}
        \caption{Variations of Z and N with pressure in the outer crust for the BSk20 and BPS models. The shaded area represent the regions covered by the Monte-Carlo procedure detailed in the text and obtained using the DZ10-GP model. See text for details.}
        \label{fig:eos:area}
    \end{figure}

It is interesting to compare the composition obtained with DZ10-GP with the predictions of other mass models, since different mass models may yield different extrapolations.
We have selected two popular mass models currently used in astrophysics: BSk20~\cite{pearson2011properties} and BPS~\cite{sharma2015unified}. The results are reported in Fig.~\ref{fig:eos:area}. The shaded area on the figure represents all the possible EoS obtained using the Monte-Carlo procedure detailed above using a 1\% cut-off on the existence probability. We observe that the results obtained with the different procedure are in good agreement with the DZ10-GP model once the error bars are properly taken into account.
It is important to notice that the transition region between the outer and inner crust is mainly governed by the mass model and not by the GP correction. As a consequence, we may expect different results using various models as shown in Fig.~\ref{fig:eos:area}.

\begin{figure}
        \centering
        \includegraphics[width=0.5\textwidth,angle=0]{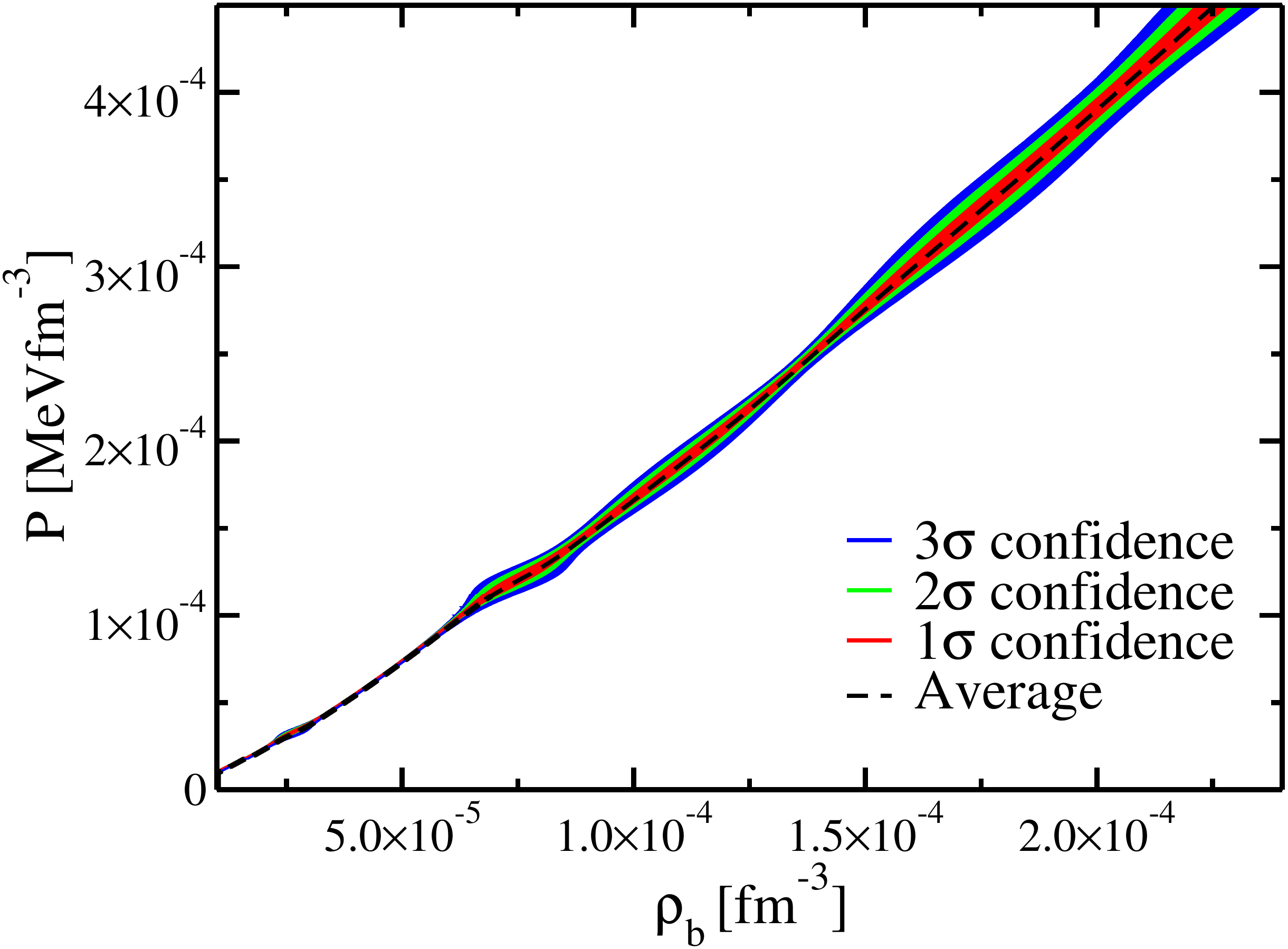}
        \caption{Equation of state, including statistical uncertainties, of the outer crust of a NS, calculated using the DZ10-GP mass model. See text for details.}
        \label{fig:eos:err}
    \end{figure}

Using the same data set, we also define a statistical uncertainty for the EoS: by counting the $10^4$ EoS built before, we define the 68\%, 95\%, and 99\% quantiles of the counts, i.e., $1\sigma$, $2\sigma$ and $3\sigma$ deviations, under the assumption that the errors follow a Gaussian distribution.
The results are presented in Fig.~\ref{fig:eos:err}. We observe that the largest uncertainties are located close to the transition from N=50 to N=82 at $P\approx1.2\times10^{-4}\text{[MeV\,fm}^{-3}]$ and approaching the transition to the inner crust at $P\approx5\times10^{-4}\text{[MeV\,fm}^{-3}]$.

%%%%%%%%%%%%%%%%%%%%%%%%%%%%%%%%%%%%%%%%%%%%%%%%%%%%%%%%%%%%%%%%%%%%%%%%%%%%%%%

\section{Conclusions}\label{sec:conclusion}

By using a Gaussian process fitted to the residuals of the Duflo-Zucker mass model, we have been able to create a mass model with a global RMS of less than $200$~keV. The resulting DZ10-GP model has the major advantage of having a very limited amount of parameters (ten in the original DZ model plus four for the GP), but it is also one of the very few mass models equipped with error bars~\cite{goriely2014uncertainties,qi2015theoretical}.
The values of the mass table are available in the Supplementary Material.

We have then applied the resulting mass model to study the composition of the outer crust of a neutron star, paying particular attention to the role of statistical errors and how they propagate to the final EoS.
Following the methodology presented in Ref.~\cite{pastore2020impact}, we have defined an existence probability of a nucleus within the crust. Such a quantity helps us to identify the possible accuracy problems related to our model, and it may help in prioritising future experimental proposals to further improve our knowledge of the crust of a neutron star.

%%%%%%%%%%%%%%%%%%%%%%%%%%%%%%%%%%%%%%%%%%%%%%%%%%%%%%%%%%%%%%%%%%%%%%%%%%%%%%%

\section*{Acknowledgements}

This work has been supported by STFC Grant No. ST/P003885/1. We also thank A. Gration for training us on the usage of Gaussian process regression.

%%%%%%%%%%%%%%%%%%%%%%%%%%%%%%%%%%%%%%%%%%%%%%%%%%%%%%%%%%%%%%%%%%%%%%%%%%%%%%%

\bibliography{biblio}

%%%%%%%%%%%%%%%%%%%%%%%%%%%%%%%%%%%%%%%%%%%%%%%%%%%%%%%%%%%%%%%%%%%%%%%%%%%%%%%

\end{document}